\documentclass[prb,twocolumn,superscriptaddress,showpacs,amssymb]{revtex4}
\usepackage[english]{babel}
\usepackage[utf8x]{inputenc}
\usepackage[T1]{fontenc}
\usepackage{amsmath}
\usepackage[dvipdfm]{graphicx}

\newcommand{\VtH}{V${}_2$H}
\DeclareMathOperator{\arccosh}{arccosh}
\DeclareMathOperator{\arcsech}{arcsech}

\begin{document}
\title{Depth-dependent ordering, two-length-scale phenomena, and crossover behavior in a crystal featuring a skin layer with defects}
\author{Charo I. \surname{Del Genio}}
	\affiliation{Department of Physics, University of Houston, 617 Science and Research 1, Houston, Texas 77204-5005, USA}
	\affiliation{Texas Center for Superconductivity, University of Houston, 202 Houston Science Center, Houston, Texas 77204-5002, USA}
\author{Kevin E. Bassler}
	\affiliation{Department of Physics, University of Houston, 617 Science and Research 1, Houston, Texas 77204-5005, USA}
	\affiliation{Texas Center for Superconductivity, University of Houston, 202 Houston Science Center, Houston, Texas 77204-5002, USA}
\author{Aleksandr L. Korzhenevskii}
	\affiliation{Institute of Problems of Mechanical Engineering, V. O. Bolshoj pr. 61, St. Petersburg 199178, Russia}
\author{Rozaliya I. Barabash}
	\affiliation{Materials Science and Technology Division, Oak Ridge National Laboratory, P.O. Box 2008, Building 4500S, 1 Bethel Valley Road, Oak Ridge, Tennessee 37831-6132, USA}
\author{Johann Trenkler}
	\affiliation{Lithography Optics Division, Carl Zeiss SMT AG, Rudolf-Eber-Straße 2, D-73447 Oberkochen, Germany}
\author{George F. Reiter}
	\affiliation{Department of Physics, University of Houston, 617 Science and Research 1, Houston, Texas 77204-5005, USA}
\author{Simon C. Moss}
	\affiliation{Department of Physics, University of Houston, 617 Science and Research 1, Houston, Texas 77204-5005, USA}
	\affiliation{Texas Center for Superconductivity, University of Houston, 202 Houston Science Center, Houston, Texas 77204-5002, USA}

\date{\today}

\begin{abstract}
Structural defects in a crystal are responsible for the ``two length-scale''
behavior, in which a sharp central peak is superimposed over a broad peak in
critical diffuse X-ray scattering. We have previously measured the scaling
behavior of the central peak by scattering from a near-surface region of a
\VtH\ crystal, which has a first-order transition in the bulk. As the temperature
is lowered toward the critical temperature, a crossover in critical behavior
is seen, with the temperature range nearest to the critical point being characterized
by mean field exponents. Near the transition, a small two-phase coexistence
region is observed. The values of transition and crossover temperatures decay
with depth. An explanation of these experimental results is here proposed by means
of a theory in which edge dislocations in the near-surface
region occur in walls oriented in the two directions normal to the surface. The
strain caused by the dislocation lines causes the ordering in the crystal to occur
as growth of roughly cylindrically shaped regions. After the regions have reached
a certain size, the crossover in the critical behavior occurs, and mean field
behavior prevails. At a still lower temperature, the rest of the material between
the cylindrical regions orders via a weak first-order transition.
\end{abstract}


\maketitle

\section{Introduction}
Since defects exist in any real system, the understanding of their influence
on ordering and structural phase transitions is important. A signature of the
presence of defects in a crystal near a phase transition is the so-called ``two
length-scale'' behavior, in which, in the critical diffuse scattering (CDS)
of X-rays or neutrons, a narrow ``central peak'' is found on top of a broad
peak~\cite{Dub79,Kor99}. Previous theoretical studies of this behavior
have established that one cause of this is the presence of dislocation
lines~\cite{Dub79,Kor99,Tre01}. These theories argue that the strain field
associated with a dislocation line results in the growth of a roughly cylindrical
ordered region near the dislocation line itself. Such regions order at a temperature
higher than the defect-free crystal. Accordingly, while the order occurs in
the cylindrical regions, the broad peak in the CDS is due to thermal fluctuations
in regions of the material which are relatively unaffected by the strain field,
while the narrow central peak is due to the fluctuations in regions where the
enhanced ordering occurs.

Unaccounted for in these theories, however, is the fact that in many real systems
defects do not exist uniformly throughout the crystal. Often defects are caused by
surface treatments or surface reconstructions and in this case they accumulate
near the surface and their density decays with depth. When this happens, the ordering
properties and two length-scale behavior depend on depth as well. Indeed, with high
resolution X-ray diffraction measurements, we have previously found that \VtH\ has
two length-scale and associated behavior that is depth dependent~\cite{Tre98,Del09}.
These measurements were performed in both reflection and transmission geometries,
allowing us to compare the behavior of the crystal at different depths.
In this paper, we propose a theoretical explanation of these experimental results
that accounts for the depth dependence of the observed behavior.

Systematic studies of many materials in which two length-scale behavior has been
found~\cite{And86,Cow96,Geh93,Geh96,Hir94,Hir95,Mcm90,Neu95,Rut97,Thu93,Wan98},
including previous studies of \VtH~\cite{Tre98}, have concluded that the narrow
central peak of the CDS only occurs in the scattering from a defective ``skin
layer'', that is a region of the material that starts a few hundred \AA{}
below the surface and extends several tens of $\mu$m below the surface. However,
to the best of our knowledge, the two-length-scale behavior in \VtH\ is different
from that which has been observed in any other material, because in \VtH\ the phase
transition in the bulk is a first-order transition. In the skin layer, instead,
the ordering is more complicated as found experimentally by a number of unusual
phenomena including: (1) diffuse scattering which, as the temperature is lowered
toward a critical value, consists of a broad peak that changes only slightly with
temperature and a narrow central peak with an amplitude that diverges~\cite{Tre98};
(2) an effective critical temperature $T_c\left(y\right)$ for the behavior of the
central peak that changes with the depth $y$ below the surface and extrapolates to
a temperature $T^{\infty}_c$ that always exceeds the bulk transition temperature~$T_0$~\cite{Del09};
(3) a crossover in the universal critical behavior shown by the central peak from
three-dimensional mean field critical behavior to a different universality class
as the temperature increases from $T_c\left(y\right)$~\cite{Tre98,Del09};
(4) a narrow two-phase region and a weak first-order transition observed
at temperatures $T_0\left(y\right)$ slightly below the critical value~\cite{Tre00}.

\begin{figure}
 \centering
\includegraphics[width=0.35\textwidth]{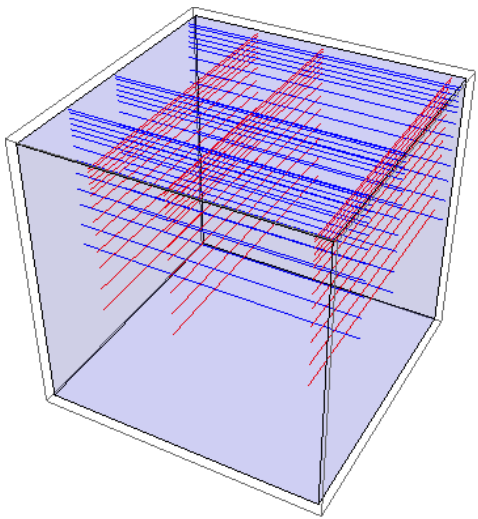}
\caption{\label{wallsketch}(Color online) Schematic illustration of the arrangement of dislocation lines in \VtH.
Edge dislocations are arranged in walls of parallel lines that extend in the directions normal
to the surface, whose density decreases with depth. Two colors are used for clarity
of presentation to distinguish the lines extending in the two directions, but they don't correspond
to any physical difference.}
\end{figure}
In order to explain these experimental findings we present a theory which
accounts for the distribution of defects experimentally detected~\cite{Tre01}: edge dislocations
occur mostly in the skin layer, accumulating near the surface; they are arranged in arrays
of parallel lines which we refer to as ``walls''; each wall consists of lines that are
oriented in either of the two directions \emph{parallel} to the surface; the walls extend
into the crystal and are thus oriented in either of the two directions \emph{perpendicular}
to the surface. In fig.~\ref{wallsketch} we show a schematic of this arrangement of defects.

\begin{figure}
 \centering
\includegraphics[width=0.45\textwidth]{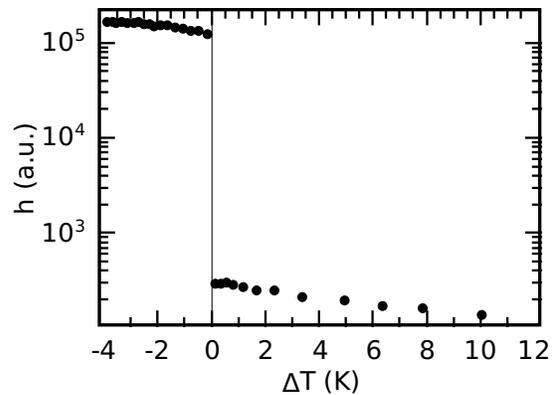}
\caption{\label{firstorder}Bulk measurements of the
peak height $h$ of the $( 0\quad\!5/2\quad\!\bar 5/2 )$ superstructure
reflection vs.\ $\Delta T=T-T_0$, where $T_0$ is the transition
temperature for the bulk. The sudden jump of a few orders of magnitude
in $h$ is a clear indication that the transition is first-order.}
\end{figure}
\begin{figure}
 \centering
\includegraphics[width=0.45\textwidth]{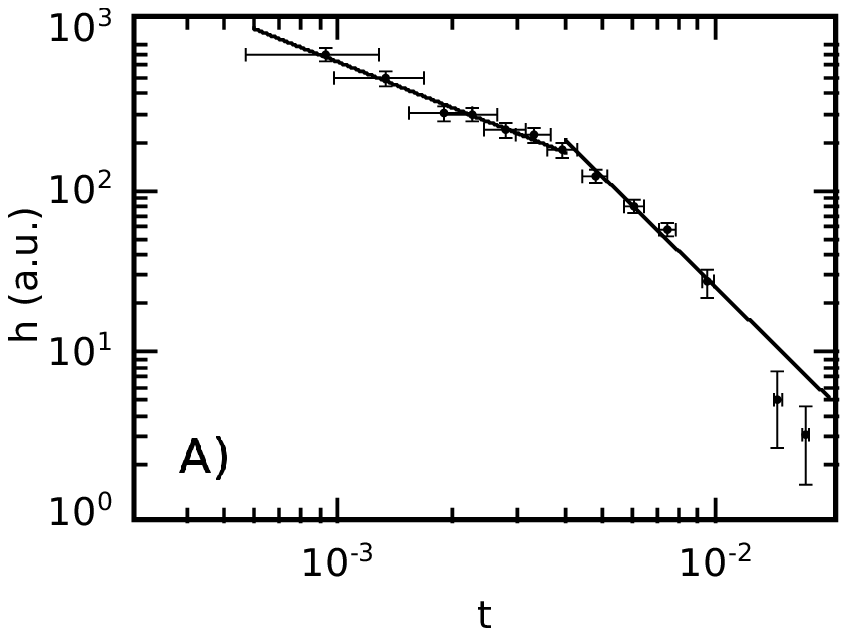}
\includegraphics[width=0.45\textwidth]{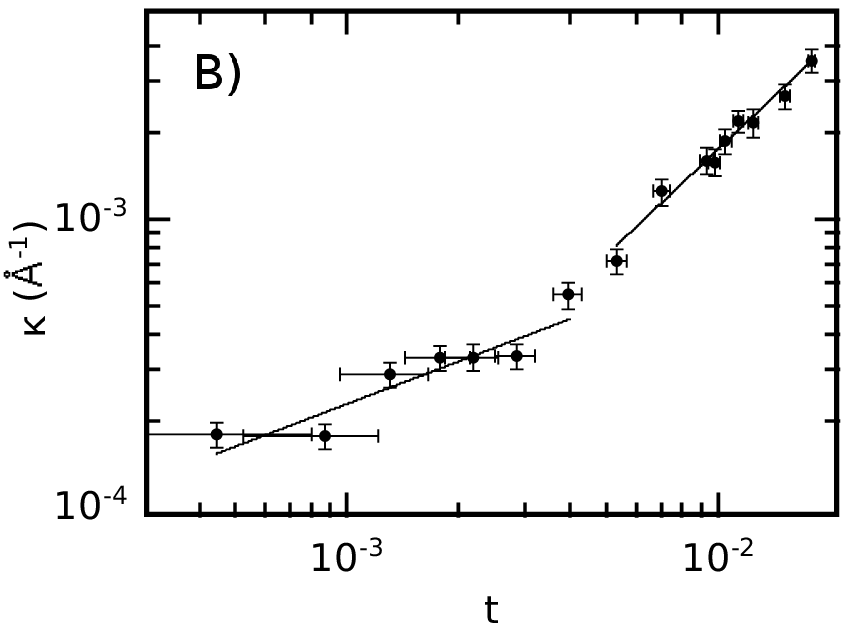}
\caption{\label{crosso}A)\ Peak height $h$ of the $( 0\quad\!5/2\quad\!\bar 5/2 )$
superstructure reflection vs.\ reduced temperature $t=\frac{T}{T_c\left(y\right)}-1$. The value
of $h$ is proportional to $t^\gamma$, and thus shows
the crossover of the critical exponent $\gamma$ from a mean-field-compatible value
of $0.93\pm0.06$ for small $t$ to $3.3\pm0.3$ for higher $t$. The measurements
were carried out at a depth $y$ of $13.1~\mu$m.\\
B)\ Inverse
correlation length $\kappa$ vs.\ reduced temperature $t=\frac{T}{T_c\left(y\right)}-1$,
showing the crossover of the critical exponent $\nu$ from a mean-field-compatible
value of $0.49\pm0.09$ for small $t$ to $1.22\pm0.09$ for higher $t$. The
measurements were carried out at a depth $y$ of $1.6~\mu$m.}
\end{figure}
\begin{figure}
 \centering
\includegraphics[width=0.36\textwidth]{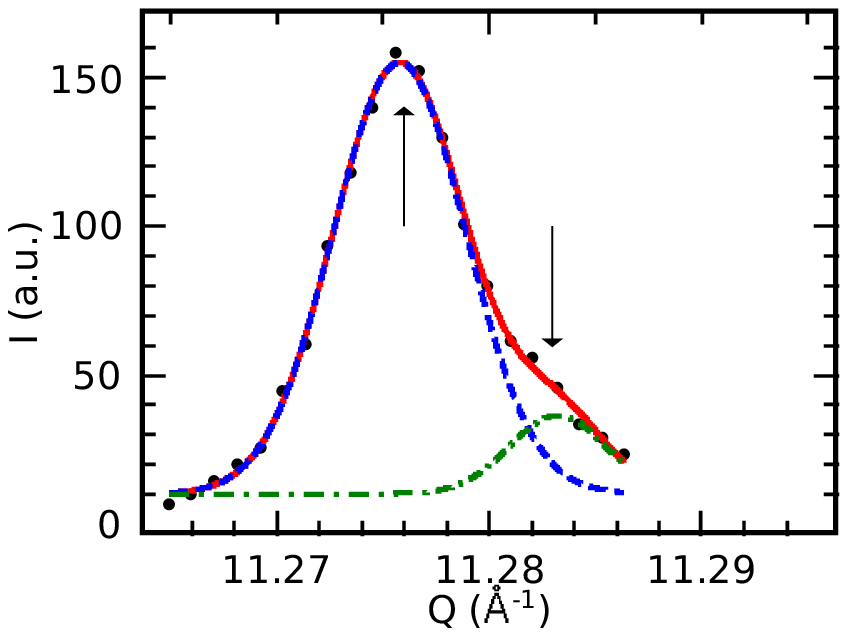}
\includegraphics[width=0.36\textwidth]{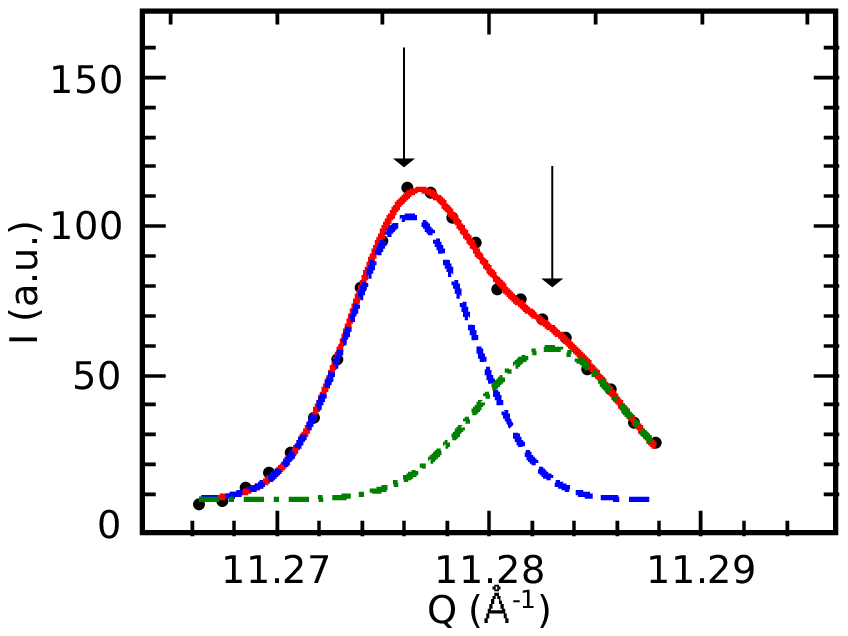}
\includegraphics[width=0.36\textwidth]{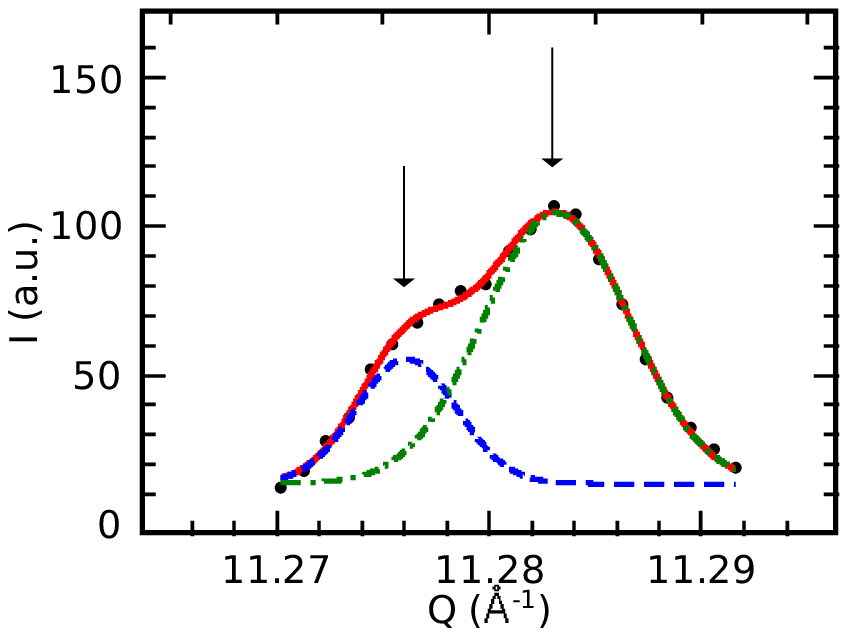}
\includegraphics[width=0.36\textwidth]{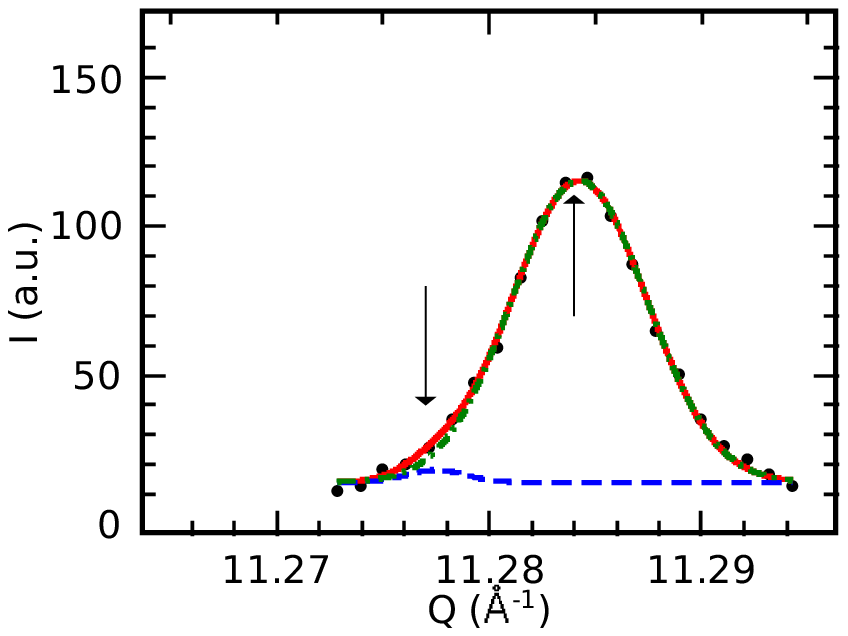}
\caption{\label{2phase}(Color online) Intensities $I$ of the $( 0\quad\!4\quad\!\bar4 )$ fundamental reflection in
a range of temperatures of $0.6$~K around the weak first-order transition temperature
${T_0}'$ in the skin-layer. From top to bottom, the measurements were taken at ${T_0}'-0.3$~K,
${T_0}'-0.1$~K, ${T_0}'+0.1$~K and ${T_0}'+0.3$~K. The dashed blue and dashed-dotted green lines
show individual gaussian peak
fits, while the red lines are the convolutions of the single peaks. At all the temperatures
we see peaks at $Q=11.276$~\AA${}^{-1}$ and $Q=11.283$~\AA${}^{-1}$ (indicated by arrows), corresponding
to the $\beta_1$ and the $\beta_2$ phases, respectively. The measurements were carried out
at a depth of $39~\mu$m.}
\end{figure}
\begin{figure}
 \centering
\includegraphics[width=0.45\textwidth]{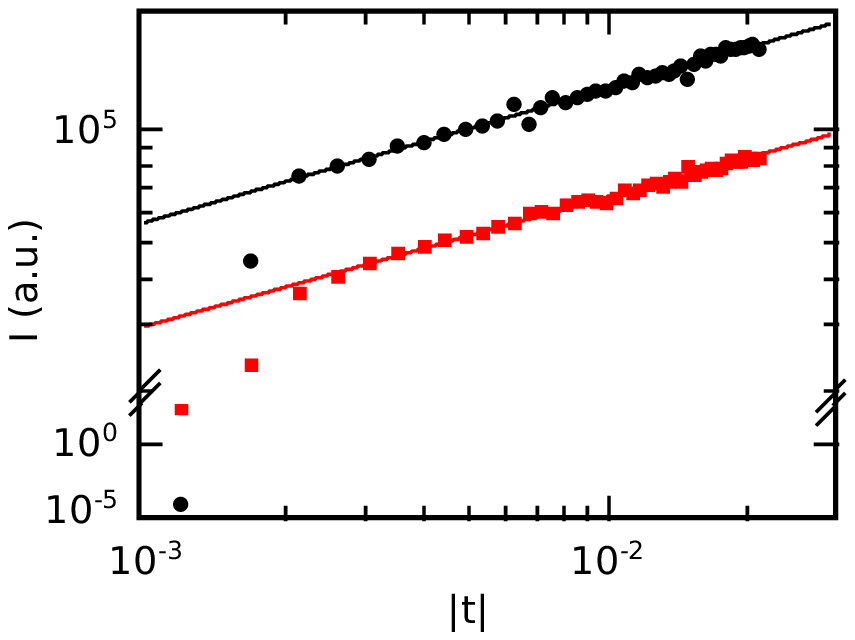}
\caption{\label{beta}(Color online) Integrated intensities $I$ vs. absolute value of reduced
temperature $t=\frac{T}{T_c\left(y\right)}-1$. The integrated intensities,
proportional to $\left|t\right|^{2\beta}$, allow us to estimate the
long range critical exponent $\beta$. The black circles correspond to the
$( 0\quad\!5/2\quad\!\bar 5/2 )$ superstructure reflection, depth $y$ of $25~\mu$m;
the red squares to the $( 0\quad\!7/2\quad\!\bar 7/2 )$ superstructure
reflection, depth $y$ of $34~\mu$m. The like-colored lines show the results
of fits which gave values for the critical exponent $\beta$ of $0.180\pm0.004$ and
$0.174\pm0.003$, respectively, obtained neglecting a narrow
($0.6$~K) region of two-phase coexistence.}
\end{figure}
As with previous theories, the strain due to dislocation lines enhances
the ordering in their vicinity. As the temperature is lowered toward the critical temperature,
ordering in the skin layer first occurs in cylindrically shaped ordered regions (ORs)
near the individual dislocation lines. However, in contrast to previous theories, which
assumed the ORs grow freely, in our theory the interaction between lines constrains
their collective growth. The ordering process responsible for the formation
of the ORs is continuous and this explains the change in the order of the phase transition
from first-order in the bulk to continuous in the skin layer. Then, as the ORs grow
and occupy more than a certain fraction of the material, the crossover in the
universal critical behavior of the central peak occurs. Since the density of dislocation
lines decreases with depth, the effective critical temperature also varies with depth.
Finally, as the temperature is reduced further, below the effective critical temperature
at any particular depth, the parts of the material in between the network
of ORs undergo a weak first-order transition. Before presenting
our theory, in the next section we recount in some detail the unusual
experimental facts of the two length-scale phenomena and associated behavior found in \VtH.

\section{Experimental results}
As in prior works~\cite{Tre01,Tre98,Del09,Tre00,Tre99}, we focus on the transition
from the ordered monoclinic $\beta_1$ phase to the disordered body centered tetragonal
$\beta_2$ phase in which the c-axis is along z (for the phase diagram see Ref.~\onlinecite{Sch79}).
In the $\beta_1$ phase, one half of the z-axis octahedral sites halfway between two
V atoms, namely the $O_{z1}$ sites, are mostly occupied by hydrogen atoms, while in
the $\beta_2$ phase both $O_{z1}$ and $O_{z2}$ sites are, on average, equally occupied~\cite{Mos83,Fuk93}.

As first reported in Ref.~\onlinecite{Tre99}, in transmission geometry there is clear
evidence of a first-order transition in the bulk material (see Fig.~\ref{firstorder}).
The diffuse scattering from the skin layer can be measured at various depths in reflection
geometry by varying beam energy and reflection order. At temperatures above the bulk
transition temperature, what we find is a broad peak corresponding to the bulk diffuse
scattering, which increases slowly with decreasing temperature, coexisting with a central peak whose
amplitude diverges at a temperature still higher than the bulk transition one~\cite{Tre98}.
The CDS of the central peak indicates that the transition is continuous in a skin layer
that has a depth of several tens of~$\mu$m. Remarkably, the temperature at which the
CDS diverges depends on the depth probed. Thus, we find a depth dependent critical temperature
$T_c\left(y\right)$, where $y$ is the effective scattering depth~\cite{Del09}. Furthermore,
as shown in Ref.~\onlinecite{Tre98}, there is a two-length scale effect in the CDS from
the skin layer. As the temperature is lowered toward $T_c\left(y\right)$, the height
of the central peak, which is proportional to the susceptibility, shows a crossover in
critical scaling behavior. In fact, for temperatures close to the critical temperature,
the value of the critical exponent $\gamma$ in the law $\chi\propto t^{-\gamma}$, describing
the divergence of the susceptibility with the reduced temperature $t=T/T_c\left(y\right)-1$,
is always consistent with 1; similarly, the full width at half maximum of the central
peak, which is proportional to the inverse correlation length $\kappa$, and which scales
as $t^\nu$, has its exponent always compatible with 0.5 (see two examples in
Fig.~\ref{crosso}). Also, we notice that, regardless of the depth, the value of $\gamma$
is always compatible with $2\nu$. Note that these values of $\gamma$ and $\nu$, measured
for small $t$, correspond to three-dimensional mean field values.

We have considered the integrated intensities $I$ of the superstructure reflections,
which are proportional to $\left|t\right|^{2\beta}$. We find that the intensities exhibit
power-law variations for temperatures below the critical value, but only
if one neglects a narrow temperature range of $0.6$~K close to the critical point, which
corresponds to a two-phase coexistence region (see Figs.~\ref{2phase} and~\ref{beta}).
The existence of this region indicates that, even in the skin-layer,
there is an actual first-order transition, albeit with a weak character,
as indicated by the power law behavior outside the aforementioned small
range of temperatures. The measured value of $\beta=0.174\pm0.003$, however, is not mean
field but it is in agreement with an earlier measurement on a different crystal~\cite{Sch87}.

Above the crossover, at larger $t$, different universal critical behavior is observed.
Measurements indicating values of $\gamma = 3.06 \pm 0.29$ and $\nu = 0.69 \pm 0.09$ have
previously been reported~\cite{Tre98}. However, other data fits at these larger values
of $t$ give a range of values for $\gamma$ and $\nu$. For example, in Fig.~\ref{crosso}
values of $\gamma = 3.3 \pm 0.3$ and $\nu = 1.22 \pm 0.09$ are found. Thus, unfortunately,
the existing experimental data sets don't allow us to determine accurately the critical
exponents for temperatures higher than the crossover. The available data show values for
$\gamma$ as low as about 1.8 and as high as 3 and values of $\nu$ from about $0.7$ to 2.

Regarding the presence of defects in the material, dislocation walls are the only kind of
defect that extends into the crystal for several $\mu$m. Other kinds of defects are not found
after the first 150~\AA{}, and the influence of this upper region on the scattering is
negligible~\cite{Del09}. The arrangement of dislocation lines into walls, in the skin
layer, is indicated by the mosaic spread. Walls occur inhomogeneously across the surface,
so that in planes parallel to the surface there is an inhomogeneous distribution of dislocation
lines~\cite{Tre01}.

\section{Theoretical model}
In order to give a picture that describes the unusual experimental
behavior of \VtH\ detailed above, let us consider a Ginzburg-Landau
free energy density expansion for a crystal with an anisotropic
distribution of dislocation lines. This free energy density is of the form
\begin{equation*}
\mathcal F\left(r\right)=\left( \nabla\eta\right) ^2+A_2\eta^2+A_4\eta^4+A_6\eta^6
\end{equation*}
where $\eta\left(r\right)$ is the order parameter field, and $A_2$, $A_4$ and $A_6$ are
coefficients that depend on position $r$, thermodynamic parameters, and, in some cases,
the strain fields created by dislocations and structural ordering. In particular,
the order parameter field is defined to be 0 for regions of the material which are in the
disordered phase, 1 for points in the ordered phase with the hydrogen atoms in the $O_{z1}$
sites, and $-1$ for the ordered phase with the hydrogen atoms in the $O_{z2}$ sites. Also
note that in the above equation the odd powers are missing as their presence is disallowed
by the symmetry of the crystal structures.

Much of the unusual phenomenological behavior observed in \VtH\ can be explained through
the behavior of the coefficients $A_2$ and $A_4$, which depends on the strain fields caused
by the edge dislocations and on the structural ordering that occurs preferentially near
them. In order to provide such explanation, we must not only understand the mechanism by
which strain fields modify $A_2$ and $A_4$ in general, but also what specific modifications
result from the particular morphological arrangement of dislocation lines that occurs in \VtH.

We will see that while the behavior of $A_2$ can explain the spatial variation
of the critical temperature, the corrections to the fourth-order term justify instead the
change in order of the transition. In particular, while a Ginzburg-Landau expansion yields
critical behavior for vanishing odd-order terms, we can still describe a first-order transition
by letting $A_4$ go negative.

In the remainder of this section, we will first discuss the effects of
the defects on $A_2$, explaining the behavior of the critical
temperature. Then, we will describe the shape
of the ordered regions near the dislocation lines. Next, we will show how the ordering
strains affect $A_4$, how this makes the transition in the skin-layer continuous, and
how the crossover occurs. Finally, we will briefly discuss the effects of the depth-dependent
crossover temperature on the size of the critical region and give an argument
for the presence of a weak first-order transition in the skin-layer.

\subsection{Critical temperature}
The compression-dilatation effect of an edge dislocation on a crystal is responsible
for a change in critical temperature~\cite{Dub79}. We can assume that the atoms of the crystal interact
more strongly where they are pushed closer to each other, and vice versa that they interact
more weakly in the opposite direction. As a first approximation we can assume that
the correction to the critical temperature is proportional to the elastic strain in
the crystal.

The elastic strain can be found following the general procedure described in Ref.~\onlinecite{Hir78}.
In the case of a single dislocation line this yields a trace of the stress tensor $\sigma$
that in polar coordinates
$r$ and $\vartheta$ centered on the dislocation line
is
\begin{equation}\label{PolarTrace}
 \mathrm{Tr}\sigma = -\frac{\mu b}{\pi}\frac{1+\nu}{1-\nu}\frac{\sin\vartheta}{r}\:,
\end{equation}
where $\mu$ is the shear modulus, $\nu$ is Poisson's ratio and $b$ is the magnitude of
the Burgers vector. Making use of the relations between elastic constants~\cite{Lan59},
we can rewrite eq.~\ref{PolarTrace} as
\begin{equation*}
 \mathrm{Tr}\sigma = -3\frac{Kb}{r}\frac{1}{2\pi}\frac{1-2\nu}{1-\nu}\sin\vartheta\:,
\end{equation*}
where $K$ is the bulk modulus. The above equation, divided by $3K$, yields the strain,
which, in our treatment, is proportional to the
local relative critical temperature change
$\tau_c$
\begin{equation}\label{NewRedLocTc}
 \tau_c\left(r\right)\propto\frac{b}{r}\frac{1}{2\pi}\frac{1-2\nu}{1-\nu}\sin\vartheta\:,
\end{equation}
where $T'_c$ is the new critical temperature, and which is defined as
\begin{equation}\label{tauC}
\tau_c=\frac{T'_c-T_c}{T_c}\:,
\end{equation}
with $T_c$ the transition temperature for an undifected crystal. Notice that this result
is in agreement with Ref.~\onlinecite{Dub79}.

In the case of multiple dislocations, the effects of the single dislocation
lines get superimposed. For walls of dislocations, this allows us to
estimate the above quantities quite easily. Without loss of generality,
we will consider a wall that extends in the $y$ direction while the lines are
parallel to $z$; we will indicate by $h\left(y\right)$ the local
inverse linear density of defects, i.e., the local average distance between
two lines. We rewrite then eq.~\ref{NewRedLocTc} in Cartesian coordinates as
\begin{equation}\label{taucart}
 \tau_c\left(r\right)\propto\frac{b}{2\pi}\frac{1-2\nu}{1-\nu}\frac{y}{x^2+y^2}\:,
\end{equation}
where we used $\sin\vartheta=\frac{y}{\sqrt{x^2+y^2}}$
and $r=\sqrt{x^2+y^2}$.
If $h\left(y\right)$ changes smoothly along the wall, we can
then write
\begin{equation}\label{walls}
\tau_c\left(r\right) \propto \frac{b}{2\pi}\frac{1-2\nu}{1-\nu}\sum_{n=-\infty}^{n=+\infty}{\frac{y+nh}{x^2+\left( y+nh\right) ^2}}\:,
\end{equation}
where $r$ is the radial distance outwards from the closest dislocation
line and the dependence of $h$ on $y$ has been omitted for simplicity of writing.
The sum of the series above is
\begin{equation}\label{series}
 \sum_{n=-\infty}^{n=+\infty}{\frac{y+nh}{x^2+\left( y+nh\right) ^2}}=\frac{\pi}{l}\frac{\sin\left( \frac{2\pi y}{h}\right) }{\cosh\left( \frac{2\pi x}{h}\right) -\cos\left( \frac{2\pi y}{h}\right) }\:,
\end{equation}
where $l$ is the unit of length used.
It should be noticed that the contribution decays exponentially with
the normal distance $x$ from the wall, in agreement with the experimental results
in Ref.~\onlinecite{Tre01}. Also, the rapid convergence of eq.~\ref{series} suggests
that $h$ needs to vary slowly only over a short distance, in order for the error
made by considering it constant to be small. Substituting eq.~\ref{series} in eq.~\ref{walls},
we can then conclude that, in the case of dislocation walls, the
local relative critical temperature change
\begin{equation}\label{CriTempWall}
 \tau_c\left(r\right)\propto\frac{b}{2l}\frac{1-2\nu}{1-\nu}\frac{\sin\left( \frac{2\pi y}{h}\right) }{\cosh\left( \frac{2\pi x}{h}\right) -\cos\left( \frac{2\pi y}{h}\right) }\:.
\end{equation}

Now, an expression can be found for the behavior of the critical temperature
with depth. In first approximation,
\begin{equation*}
 T'_c\left(y+h\right)\approx T'_c\left(y\right)+\frac{\partial T'_c\left(r\right)}{\partial y}\delta y\:,
\end{equation*}
which can be rewritten as
\begin{equation*}
 T'_c\left(y+h\right)\approx T'_c\left(y\right)+\chi\:,
\end{equation*}
where
\begin{equation*}
 \chi\propto T_ch\frac{\partial\tau_c\left(r\right)}{\partial h}\frac{\mathrm dh}{\mathrm dy}\:.
\end{equation*}
Note that $\frac{\mathrm dh}{\mathrm dy}$ can be computed from experimental data, such
as the ones reported in Fig.~5 of Ref.~\onlinecite{Tre01}, while the other derivative
can be computed from Eq.~\ref{CriTempWall}, yielding the following expression for $\chi$:
\begin{widetext}
\begin{equation*}
 \chi\propto T_c\frac{b}{2l}\frac{1-2\nu}{1-\nu}\frac{\mathrm dh}{\mathrm dy}\left(\frac{h\sin\left(\frac{2\pi y}{h}\right)\left[x\sinh\left(\frac{2\pi x}{h}\right)+y\sin\left(\frac{2\pi y}{h}\right)\right]}{{\left\lbrace x\left[\cosh\left(\frac{2\pi x}{h}\right)-\cos\left(\frac{2\pi y}{h}\right)\right]\right\rbrace}^2}-\frac{y\cos\left(\frac{2\pi y}{h}\right)}{h\left[\cosh\left(\frac{2\pi x}{h}\right)-\cosh\left(\frac{2\pi y}{h}\right)\right]}\right)\:.
\end{equation*}
\end{widetext}
Notice that using typical values for the elastic constants, the proportionality
factors in front of the functions in the above equations are of the order of
unity.

As for $A_2$, the second order coefficient in the free energy density expansion,
we recall that without defects it is $A_2=a\left(T-T_0\right)$, with $a$ generic
proportionality constant. From this expression and the definition of $\tau_c$,
we can write the new second order coefficient of the expansion as
\begin{equation*}
  A'_2\left( r\right) = \left(1+\tau_c\right)A_2-aT\tau_c\:.
\end{equation*}
The spatial dependence of the parameter is contained in the
local relative critical temperature change,
that we wrote here as $\tau_c$ for sake of simplicity, but which
is actually $\tau_c\left( r\right)$. Thus, in contrast to $A_2$, $A'_2$ varies
in space.

Furthermore, summing the contributions coming from single dislocations or dislocation
walls, one gets a field of
relative critical temperature changes,
which features in the free
energy density expansion as follows:
\begin{equation*}
\mathcal F\left( r\right) = \left( \nabla\eta\right) ^2+a\lbrace T-T_0\left[ \tau_c\left( r\right) +1\right] \rbrace \eta^2+A_4\eta^4+A_6\eta^6\:.
\end{equation*}

In this treatment the strain field surrounding a dislocation line is spatially
inhomogeneous, and this leads to a spatial inhomogeneity of the ordering near
a dislocation line. In fact, the strain field caused by a line is dipole-like.
Because of this, at temperatures above the transition temperature of an undefected
crystal, ordered regions exist below the dislocation lines, where the transition
temperature is increased. On the other hand, the transition temperature on the
other side of the line is symmetrically decreased. Thus, disordered regions exist
within a dislocation wall even at a temperature $T$ below the transition
temperature of an ideal crystal. Then, at any depth, the correlation
length along $y$ cannot exceed a value of the order of $h/2$; conversely, the
local ordering temperature can only change when the density of defects changes,
i.e., over a distance of at least a few $h$.

\subsection{Shape of the ordered regions}
To find the shape of the ordered regions, we first fix a value for the
relative critical temperature change
$\tau_c$. Then, we solve eq.~\ref{CriTempWall} in order to
find expressions relating $x$ and $y$ as functions of each other. These solutions
correspond to the border of the ordered region for the chosen value of $\tau_c$,
and, up to a multiplicative factor
that is of order of unity for typical values
of the elastic constants,
they are
\begin{widetext}
\begin{gather}\label{yofx}
y\left(x\right)=\frac{h}{\pi}\arctan\left\lbrace\frac{\pi\pm\sqrt{\pi^2+\tau_c^2\left[1-\cosh^2\left(\frac{2\pi x}{h}\right)\right]}}{\tau_c\left[\cosh\left(\frac{2\pi x}{h}\right)+1\right]}\right\rbrace\:;\\
x\left(y\right)=\pm\frac{h}{2\pi}\arccosh\left\lbrace\frac{2\pi\tan\left(\frac{\pi y}{h}\right)+\tau_c\left[1-\tan^2\left(\frac{\pi y}{h}\right)\right]}{\tau_c\left[1+\tan^2\left(\frac{\pi y}{h}\right)\right]}\right\rbrace\:.\label{xofy}
\end{gather}
\end{widetext}
Note that the regions are symmetric with respect to the $x$ and $y$ directions.

From eq.~\ref{yofx} it follows that the maximum extension of the regions along
the $y$ direction, $D_y$, happens at $x=0$, and its value is
\begin{equation}\label{minor2}
D_y=\frac{h}{\pi}\arctan\left(\frac{\pi}{\tau_c}\right)\:.
\end{equation}
Notice that $D_y\rightarrow\frac{h}{2}$ when $\tau_c\rightarrow0$, that is, the
regions can never extend in $y$ for a distance greater than $\frac{h}{2}$, as
already stated in the previous section.

From eq.~\ref{xofy}, it follows that the maximum estension along the $x$ direction,
$D_x$, happens at $y=\frac{D_y}{2}$, and its value is
\begin{widetext}
\begin{equation*}
D_x=\frac{h}{\pi}\arccosh\left(\frac{2\pi\tan\left[\frac{1}{2}\arctan\left(\frac{\pi}{\tau_c}\right)\right]+\tau_c\left\lbrace1-\tan^2\left[\frac{1}{2}\arctan\left(\frac{\pi}{\tau_c}\right)\right]\right\rbrace}{\tau_c\left\lbrace1+\tan^2\left[\frac{1}{2}\arctan\left(\frac{\pi}{\tau_c}\right)\right]\right\rbrace}\right)\:,
\end{equation*}
\end{widetext}
that can be rewritten as
\begin{equation}\label{major2}
D_x=\frac{h}{\pi}\arcsech\left(\frac{\tau_c}{\sqrt{\tau_c^2+\pi^2}}\right)\:.
\end{equation}
Note that the center of the region is at $x=0, y=\frac{D_y}{2}$.

Now it is possible to show that the ordered regions grow from a rod-shaped nucleus
with a circular cross-section. With growth, their aspect ratio changes, making the cross-section
effectively elliptical. Finally, the constraint constituted by the order-inhibited zones
above the dislocation lines introduces distortions in the shape. To show this, we prove
that, for large values of $\tau_c$, corresponding to the beginning of the growth of the
regions, $\frac{x^2}{a^2}+\frac{y_C^2}{b^2}\approx1$, where $a=\frac{D_x}{2}$,
$b=\frac{D_y}{2}$ and $y_C=y-b$ is the $y$ coordinate measured from the center of the
region. In this regime, we can expand $b$ as follows:
\begin{equation*}
b\approx\frac{h}{2\tau_c}+O\left(\tau_c^{-3}\right)\:.
\end{equation*}
Then, we use a similar treatment for $a$. First we expand the argument of the $\arcsech$ for
large values of $\tau_c$:
\begin{equation*}
a\approx\frac{h}{2\pi}\arcsech\left[1-\frac{\pi^2}{2\tau_c^2}+O\left(\tau_c^{-4}\right)\right]\:.
\end{equation*}
Then, we expand the $\arcsech$ for values of the argument close to 1, getting
\begin{equation*}
a \approx\frac{h}{2\tau_c}+O\left(\tau_c^{-3}\right)\:.
\end{equation*}

Expanding the argument of the $\arctan$ in eq.~\ref{yofx} for large $\tau_c$ we get
\begin{equation*}
y \approx \frac{h}{\pi}\arctan\left[\frac{\pi}{\tau_c}\frac{1}{\cosh\left(\frac{2\pi x}{h}\right)+1}+O\left(\tau_c^{-2}\right)\right]\:,
\end{equation*}
where we have imposed the condition that the argument be real. Note that a big $\tau_c$
implies a small $x$. Then, it is
\begin{equation}\label{nucy}
y\approx\frac{h}{2\tau_c}+O\left(\tau_c^{-3}\right)\:.
\end{equation}

\begin{figure}
 \centering
\includegraphics[width=0.45\textwidth]{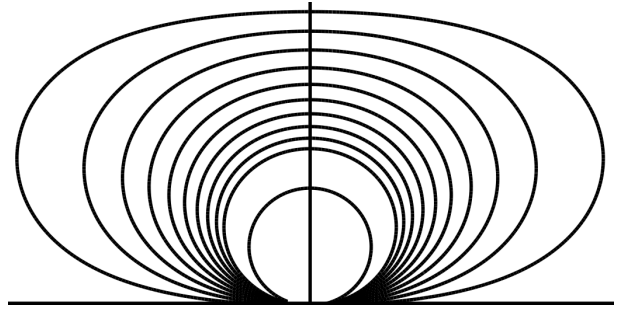}
\caption{\label{CylReg}Shape of the cross section of the ordered regions
for different values of $\tau_c$. From the outermost line inwards, the lines
correspond to values of $\tau_c$ that range from $0.1$ to 1, in steps of $0.1$,
except for the innermost line that corresponds to $\tau_c=1.5$.}
\end{figure}
Also note that in the beginning of the growth $a=b$. This means that the regions
nucleate in the shape of narrow cylindrical rods along the dislocation
lines. Figure~\ref{CylReg} shows the shape of the cross section of the
ordered regions for different values of $\tau_c$.

Substituting eq.~\ref{nucy} into eq.~\ref{xofy}, and expanding for high values of $\tau_c$ we find:
\begin{equation*}
x \approx\pm\frac{h}{2\tau_c}O\left(\tau_c^{-3}\right)\:.
\end{equation*}

Finally, we have
\begin{equation*}
\lim_{\tau_c\rightarrow\infty}\frac{x^2}{a^2}+\frac{y_C^2}{b^2}=\frac{h^2}{4\tau_c^2}\frac{4\tau_c^2}{h^2}+\left(\frac{h}{2\tau_c}-\frac{h}{2\tau_c}\right)^2\frac{4\tau_c^2}{h^2}=1\:,
\end{equation*}
that is, for small sizes, the regions are effectively elliptical,
and $a$ and $b$ can be identified with the major and minor semiaxes,
respectively.

\begin{figure}
 \centering
\includegraphics[width=0.3\textwidth]{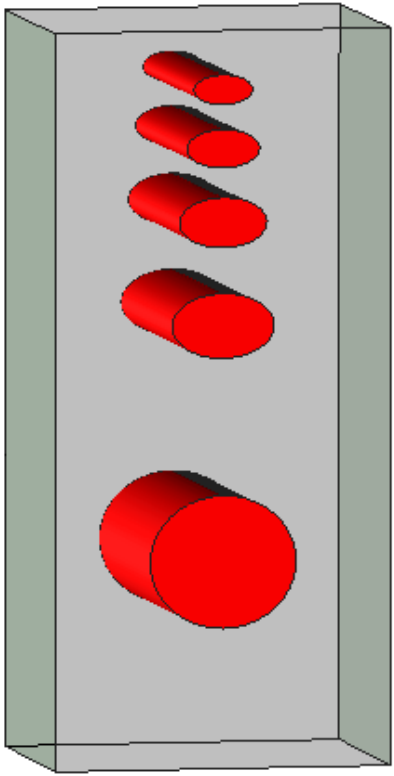}
\caption{\label{dislwall}(Color online) Schematic illustration of the shape and size of the
ordered regions in a wall at a fixed temperature. The surface of the crystal
is at the top of the image. The crystal extends beyond what pictured, that is
just a region in the vicinity of a single wall. While in the skin-layer the regions
have an elliptical cross-section whose aspect ratio is constant, at greater
depths their shape becomes increasingly circular. Also, while the size of the regions
increases with depth, their number density decreases.}
\end{figure}
To understand the relative shape and size of the ordered regions in a wall
at a particular fixed temperature, first notice that from the definition
of $\tau_c$ (eq.~\ref{tauC}) it follows that any given temperature $T$ is critical
for points in the crystal that satisfy the equation
\begin{equation*}
T=T_c\left(\tau_c+1\right)\:,
\end{equation*}
that is to say, for points in which $\tau_c=\frac{T-T_c}{T_c}$.
Then, from eqs.~\ref{minor2} and~\ref{major2}, for a fixed value of
$\tau_c$ the major and minor semiaxes of the ordered regions grow linearly with
$h$. Consequently, for a fixed temperature, the density of defects decreasing
with depth makes the size of the ordered regions grow with depth. Their aspect
ratio, in contrast, remains the same. Of course, this argument is valid as long
as the approximation employed to obtain eq.~\ref{CriTempWall} using eq.~\ref{series}
holds. Deep in the skin-layer, where the density of defects almost vanishes,
the strain field associated with any ordered region is restricted to that if a
single dislocation line. Therefore, at great depths, the expression for $\tau_c$
tends to the form given in eq.~\ref{taucart}. Solving eq.~\ref{taucart} yields
\begin{gather*}
y\left(x\right)=\frac{1\pm\sqrt{1-16\pi^2x^2\tau_c^2}}{4\pi\tau_c}\\
x\left(y\right)=\pm\sqrt{\frac{y\left(1-2\pi y\tau_c\right)}{2\pi\tau_c}}\:.
\end{gather*}
We can then calculate $D_y$ and $D_x$ as above:
\begin{gather*}
D_y = y\left(0\right)=\frac{1}{2\pi\tau_c}\\
D_x = 2x\left(\frac{D_y}{2}\right)=\frac{1}{2\pi\tau_c}\:.
\end{gather*}
Thus, as the depth increases, the shape of the cross-section of the
regions becomes increasingly circular (see fig.~\ref{dislwall}).

\subsection{Critical behavior}
To determine how the defects affect the fourth-order term in the free energy density
expansion, we start with finding the displacement field due to
a point source of expansion by means of the tensor Green's function for
the equilibrium equation. The displacement, which is purely radial, is
\begin{equation}\label{raddispl}
 \upsilon_r=\frac{W}{4\pi\left( \lambda+2\mu\right) }\frac{1}{r^2}\:,
\end{equation}
where $W$ is the work associated with this expansion, $\lambda$ is
Lamé's constant, and $\mu$ is the shear modulus~\cite{Hir78}.

Then, consider the particular case of a local expansion due to the difference of structure
between 2 points. Given our order parameter field $\eta$, this expansion, purely hydrostatical
since it's applied to an infinitesimal volume, can be thought of as due to an equivalent pressure
\begin{equation*}
 p=K\frac{\delta V}{V}\left[ \eta\left( \vec{r}\right) -\eta\left( \vec{r'}\right) \right] =K\frac{\delta V}{V}\eta\left( \vec{r}\right)\:,
\end{equation*}
where $K$ is the bulk modulus. Since we are associating a change of volume with a point,
what we want to consider is the smallest volume for which it makes sense to think about
a change of structure, that is, the volume of a unit cell. So, calling $v_O$ the volume
of a unit cell in the ordered phase and $v_D$ the one in the disordered phase,
we have $\delta V=v_O-v_D$, hence
\begin{equation}\label{Pressure}
 p=K\left( \frac{v_O}{v_D}-1\right) \eta\left( \vec{r}\right) \:.
\end{equation}
The work done to achieve this deformation is
\begin{equation*}
  W=p\delta V=K\frac{\left( v_O-v_D\right) ^2}{v_D}\eta\left( \vec{r}\right)\:,
\end{equation*}
and we can use this expression in eq.~\ref{raddispl} to find the contribution
to the displacement at the point $\vec{r}$ due to a point at $\vec{r'}$, which is
\begin{equation*}
 \upsilon_r=\frac{K}{4\pi\left( \lambda+2\mu\right) }\frac{\left[ v_O-v_D\left( \vec{r'}\right) \right] ^2}{v_D\left( \vec{r'}\right) }\frac{\eta\left( \vec{r}\right) }{r^2}\:.
\end{equation*}
Notice that we have made the dependence of $v_D$ on the point explicit. For any
particular $\vec{r}$, $v_O-v_D\left( \vec{r}\right) $ vanishes in the limit of
$T\rightarrow T_c\left( {\vec{r}}\right) $.
Then, knowing the field in every point, the total displacement at $\vec{r}$ is
\begin{equation}\label{TotalDispl}
\begin{split}
 \vec{u}\left( \vec{r}\right) & = \frac{K}{4\pi\left( \lambda+2\mu\right) }\eta\left( \vec{r}\right) \\
& \quad\times\int\mathrm{d}V'\frac{\left[ v_O-v_D\left( \vec{r'}\right) \right] ^2}{v_D\left( \vec{r'}\right) }\frac{\vec{r'}}{{r'}^3}\:,
\end{split}
\end{equation}
where the integral is extended over all the points in the disordered phase.

Strictly speaking, eq.~\ref{Pressure} is valid when the transition is from the
disordered to the ordered phase. For the inverse transition, $\delta V$ has the
opposite sign and the term in parentheses in eq.~\ref{Pressure} is $\frac{v_D}{v_O}-1$. Yet, since
$v_D\approx v_O$, the error committed in using the same formula is very small; in
fact, in the subsequent equations the volume difference ends up squared, so that
the error is actually of second order and is neglected in this linear treatment.
Notice also that, after integration, the displacement is not necessarily purely hydrostatic.

From the displacement vector we can find the components of the strain tensor
in Cartesian coordinates as
\begin{equation}\label{DisplTens}
 u_{ij}=\frac{1}{2}\left( \frac{\partial u_i}{\partial x_j}+\frac{\partial u_j}{\partial x_i}\right) \:.
\end{equation}
Then, from the generalized Hooke's law, we can find the correction to the
free energy density due to the deformation induced by the phase transition, $B$, as
\begin{equation}\label{A4Corr}
 B=\mu\left( u_{ij}-\frac{1}{3}\delta_{ij}u_{kk}\right) ^2+\frac{1}{2}Ku_{kk}^2\:.
\end{equation}
It's to be stressed that this quadratic form is always positive,
since $K$ and $\mu$ are always positive. In fact, since the body is in equilibrium,
in the absence of forces the energy must have a minimum at $u_{ij}=0$, that is, the quadratic
form must be positive. Since this has to happen in any case, it also has to happen when the
stress is a pure shear or a pure hydrostatic compression, that is when only one of the two
addenda is non-zero. This is the reason why both $K$ and $\mu$ must always be positive~\cite{Lan59}.
Moreover the squares make this correction
actually proportional to $\eta^2$. Then, when it is applied to the $2^{\mathrm{nd}}$~order
coefficient in the free energy density expansion, where it belongs since its functional form is that
of a strain, it yields a positive effective correction to the $4^{\mathrm{th}}$~order term.

This correction, which corresponds exactly to the elastic free energy,
does not depend on the actual structure of defects that causes
it, but merely on the order parameter field and, of course, on the elastic constants
of the material. This means that, regardless if it's coming from a single dislocation,
a wall of dislocations, or a more complicated framework of defects, once the
elements of the strain tensor are known, they can be used to correct directly
the free energy density expansion as discussed and as follows:
\begin{equation*}
\begin{split}
 \mathcal F\left( r\right) & =\left( \nabla\eta\right) ^2+a\lbrace T-T_0\left[ \tau_c\left( r\right) +1\right] \rbrace \eta^2\\
& \quad+\left[ A_4+\frac{B}{\eta^2}\right] \eta^4+A_6\eta^6\:,
\end{split}
\end{equation*}
where $\tau_c$ is the
local relative critical temperature change
resulting
from the superposition of the contributions of single dislocation
lines (eq.~\ref{NewRedLocTc}) and dislocation walls (eq.~\ref{CriTempWall}).
While no quantitative solution of the above equations is offered here,
notice that an explicit calculation would be very difficult. In fact, even in
simpler cases of lattices in only 2 dimensions featuring regularly arranged
point defects with short range interactions, the results can be highly
non-trivial~\cite{Fis75}.

We know, however, that, from a purely phenomenological point of view,
the change of order in a phase transition can be associated
with the change of the sign of $A_4$, which determines whether the transition
is continuous or discontinuous. The fact that the correction to this coefficient
is always positive, shows how an originally negative $A_4$ can turn
positive, thus resulting in a first-order transition becoming continuous
in the presence of defects, as observed in \VtH.

Note that a correction to $A_4$ due to the ordering process exists also in an
undefected material. However, at least in the beginning of the nucleation of
the new phase in the disordered material, this correction is not greater than
the one in the skin-layer discussed above\cite{Bru81}. In fact, when the ordered
phase starts to nucleate in a perfect crystal, the shape of the nucleus is that
of an infinitely thin platelet~\cite{Kha83}. The strain energy density contribution
of the infinitesimal platelet is
\begin{equation}\label{A4b}
\mathcal F_p=\frac{1}{2}C\left(\hat n_0\right)\:,
\end{equation}
where $C$ is a linear function of the strains and $\hat n_0$ is the unitary
vector minimizing $C$~\cite{Kha83}. As before, the strains being proportional
to $\eta^2$, the equation above yields an effective correction on the
fourth-order term.

On the other hand, we have seen that the shape of the nucleus in the skin-layer
is that of an infinitesimal cylindrical rod. Then, eq.~\ref{A4b} becomes
\begin{equation*}
\mathcal F_s=\frac{1}{2}\left\langle C\left(\hat{n}\right)\right\rangle\:,
\end{equation*}
where the brackets indicate the angular average of $C$. Since the
average of a quantity is never smaller than its minimum, the correction
in the skin layer due to an infinitesimal nucleus cannot be smaller than the
one in the bulk.

\subsection{Tricritical behavior}
\begin{figure}
 \centering
\includegraphics{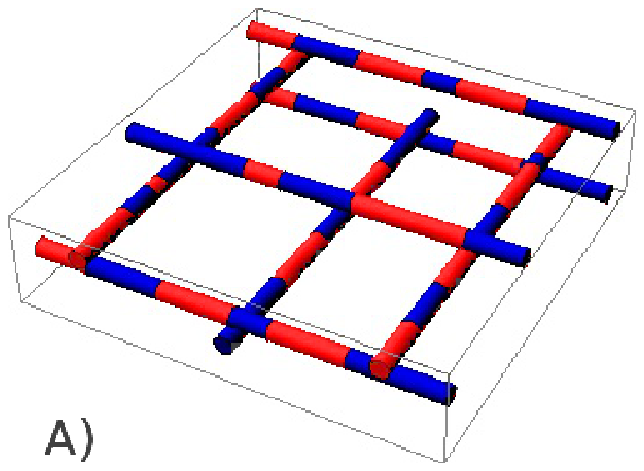}
\includegraphics{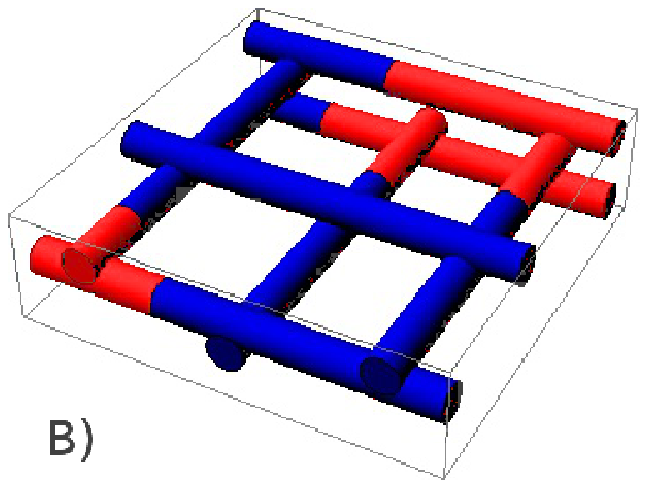}
\caption{\label{cylreg}(Color online)
Schematic illustration of the ordering process near dislocation lines
at an arbitrary depth.
Red and light blue regions of the ordered regions distinguish different
ordering domains, corresponding to the two different signs of the order
parameter. At higher temperatures, shown in A), the regions don't 
touch each other, the correlation length (average length of an ordered
domain) is smaller than the distance between dislocation walls, and there
is little or no correlation in the ordering in different regions.
At lower temperatures, shown in B), the ordered regions are thicker
and now touch, the correlation length is longer than the average distance
between dislocation walls, and the order is spreading throughout the
network of regions.}
\end{figure}
Recalling eq.~\ref{TotalDispl}, we notice that the more
the order spreads, the more the domain over which the intergral is carried out
is reduced. Moreover, the contributions in the integral are weighted with the
square of the distance to the point to which they apply.

Thus, with growing regions, the magnitude of the displacement vector will steadily decrease.
Also, the displacement vector field will become spatially increasingly
homogeneous, especially in points not too close to the borders.
Therefore, the components of the strain tensor will decrease as well,
since we know from eq.~\ref{DisplTens} that they
depend on the derivatives of the total displacement vector components,
which are changing smoothly.

The fact that the correction to $A_4$ is
proportional to the square of a linear combination of strain tensor components
(eq.~\ref{A4Corr}) guarantees that its value will be small and positive
throughout the new transition.
If this is added to a small negative ``original'' fourth~order coefficient,
as it is the case in \VtH, whose bulk transition is first order, it will
make $A_4$ effectively vanish, thereby making the
transition behavior tricritical, and, because the upper critical dimension
of this tricritical theory is 3, causing the mean-field values of the exponents
that are observed experimentally.

Note that below a certain temperature where the wall lines have
ordered independently, the walls consist of domains with order
parameter of opposite signs. In the case of a complex framework of
dislocation walls which intersect and interact with each other,
on further lowering the temperature, one of the two signs will
eventually spread throughout the network until the order parameter
has the same sign everywhere (see also Fig.~\ref{cylreg}).

\subsection{Effects on the critical region}
The mechanism for tricriticality also provides an explanation for the depth-dependence
of the crossover temperature. In fact, the density of defects decreases with depth.
Thus, in order for the domain of the integral in eq.~\ref{TotalDispl} to be sufficiently
reduced, the temperature needs to be decreased more at greater depths.
Consequently, the correlation length $\xi$ has to increase more at greater depths than it has
to in layers closer to the surface in order to reach the critical point. In other words,
for the same change in reduced temperature, the change in correlation length is higher
for shallower depths. The corresponding scaling law should then be modified so that the
coefficient depend on depth:
\begin{equation*}
 \xi(d)=a(d)\,t^{-\nu}\:.
\end{equation*}
In particular, it is reasonable to expect $a(d)$ to be proportional to the density of
defects, as it is in fact observed experimentally~\cite{Del09}.
This mechanism provides an explanation as to why experimental measurements of $\xi$ at
different depths yield different values even if taken at corresponding reduced temperatures,
once the depth-dependence of the critical temperature has been accounted for, and even
though the critical exponent remains the same. A treatment to collapse the data onto a single
curve, including corrections for the experimental method used, has been proposed and verified
in Ref.~\onlinecite{Del09}.

\subsection{Weak first-order transition}
As the temperature is lowered below the critical point, the remaining material
between the walls orders and a second, distinct, phase transition occurs.
However, considering eq.~\ref{TotalDispl} again, we notice
that the correction only applies to the material inside the
``skeleton'' of ORs. Therefore, for the material that is still between the ORs,
the fourth-order coefficient is still negative. In such regions the transition
is still first-order, but it is weaker than it is in an undefected crystal.

Because this transition is only weakly first-order,
a critical exponent $\beta$ associated with it can still be measured as shown in Fig.~\ref{beta}.
The measured values are $0.180\pm0.004$ and $0.174\pm0.003$.
Such values confirm that the transition regime is not
tricritical, since in that case one would expect a mean field exponent $\beta=0.25$~\cite{Nel75}.

\section{Conclusions}
In conclusion, we have proposed a theoretical model which explains the 
two-length-scale phenomena and related behavior observed experimentally in many 
materials, including the unusual
ordering behavior observed in \VtH. In particular, the depth-dependence
of the critical temperature in the skin layer, reported recently \cite{Del09},
is shown to be caused by the strain field
induced by the presence of walls of dislocations. The ordering process itself crosses over
between two regimes, causing the experimentally observed crossover in the
values of the critical exponents.
The additional strains induced by the formation of the ordered 
regions are responsible for increasing
the effective value of the fourth-order coefficient in a 
Ginzburg-Landau free energy density expansion, thereby
allowing the transition to become continuous in the skin layer
even if it is first-order in the bulk as we find in \VtH. 
Furthermore, the strength of this correction weakens during
the spreading of the order through the network, 
driving the value of the fourth-order coefficient to
zero, thus producing
tricritical behavior and mean-field values of the critical exponents.
Thus, when the temperature is lowered, first a continuous
transition happens along the dislocations as described. Then,
at a still lower temperature, the material between the cylindrical regions
orders. However, it undergoes a transition that is first order,
similar to the one that takes place in the bulk but weaker.
The model also explains why, in the mean-field regime, 
the critical region has a depth-dependent size.

Although in this paper we have specifically considered the case
of \VtH, the theoretical framework that we have developed should
be broadly applicable for understanding
ordering behavior in defective materials, particularly those
that have an anisotropic distribution of defects. We considered
a situation in which the bulk, and the material in between the 
interconnected network of cylindrical regions in the skin layer, 
orders through first-order
transitions. However, the model we have developed is easily
adaptable to a situation where the transition in between the
cylindrical regions is continuous, and to a situation where
both transitions are continuous. In the latter case, though,
we would not expect to see a crossover in the scaling behavior
of the central peak of the skin layer CDS.

\begin{acknowledgments}
This work is supported by the NSF through grants~\#DMR-0406323 and~\#DMR-0908286
(KEB and CIDG). KEB, CIDG and SCM also acknowledge support by the Texas Center
for Superconductivity at the University of Houston (T${}_\mathrm{c}$SUH).
RIB is supported by the Division of Materials Science and Engineering,
Office of Basic Energy Science, U.S. Department of Energy.
We gratefully acknowledge M.~E.~Fisher for his valuable comments and discussions.
\end{acknowledgments}

\end{document}